\documentclass[fleqn,twoside]{article}
\usepackage{espcrc2}
\usepackage{epsfig,graphicx}
\usepackage[figuresright]{rotating}
\newcommand{\e}{\mathrm{e}}
\newcommand{\im}{\mathrm{i}}
\newcommand{\Tr}{\mathrm{Tr}}
\renewcommand{\Re}{\mathrm{Re}}

\title{Lattice QCD at finite temperature: Evidence for calorons from
  the eigenvectors of the Dirac operator\thanks{talk given by
W. S\"oldner at Lattice 2001, Berlin}}

\author{C. Gattringer\address[home]{Institut f\"ur Theoretische Physik,
Universit\"at Regensburg, D-93040 Regensburg, Germany},
M. G\"ockeler\addressmark[home], P.E.L. Rakow\addressmark[home],
A. Sch\"afer\addressmark[home], W.  S\"oldner\addressmark[home], and
T. Wettig\address{Center for Theoretical Physics, Yale University, New
Haven, CT 06520-8120, USA}$^,$\address[riken]{RIKEN BNL Research Center,
Brookhaven National Laboratory, Upton, NY 11973-5000, USA}}
 
\begin{document}  
   
\begin{abstract}
  We analyze the eigenvalues and eigenvectors of the staggered Dirac
  operator in quenched lattice QCD in the vicinity of the
  deconfinement phase transition using the L\"uscher-Weisz gauge
  action. The spectral and localization properties of the low-lying
  eigenmodes show characteristic differences between the $Z_3$ sectors
  above the critical temperature $T_c$.  These findings can be
  interpreted in terms of calorons.
\end{abstract}

\maketitle

\section{INTRODUCTION}
The chiral phase transition is a main feature of QCD. In the instanton
picture \cite{Sha1} the chirally broken phase is represented by an
ensemble of weakly interacting instantons and anti-instantons. Due to
the fermionic quasi zero modes of the (anti-) instantons, a finite
density of eigenvalues, $\rho_{\rm Dirac}$, is generated near zero in
a (large) volume $V$, and according to the Banks-Casher
formula~\cite{BC}
\begin{equation}
\langle \bar q q \rangle = -\,{\pi\rho_{\rm Dirac}(0)/ V}
 \label{BCrel} 
\end{equation}
chiral symmetry is broken. In the chirally symmetric phase instantons
and anti-instantons form strongly bound molecules, the eigenvalues of
the Dirac operator move towards the bulk of the spectrum and the
eigenvalue density vanishes near the origin. Of course one would like
to verify the essential parts of this picture in QCD. Therefore we
study the localization properties of the eigenmodes of the lattice
Dirac operator in the neighborhood of the chiral phase transition. In
a previous paper
\cite{own} we have investigated this question with the Wilson gauge
action on smaller lattices. Here we present results obtained on larger
lattices with the L\"uscher-Weisz gauge action.

\section{PROPERTIES OF CALORONS}
The temporal extension $L_t$ of the lattice is related to the
temperature $T$ by $aT = 1/L_t$, where $a$ is the lattice spacing. Due
to the periodic boundary condition in the time direction, an instanton
becomes a chain of instantons which one calls caloron
\cite{Sha1,shepard}. The properties of a caloron are similar to that
of an instanton: a caloron has topological charge one and a localized
left-handed fermionic zero mode. However, the localization
pattern of the zero mode is different. Whereas in the case of an
instanton the zero mode is localized in space and time, the zero mode
of a caloron is localized in space but delocalized in time.

Let us now compare the localization of the zero mode in the different
sectors of the Polyakov loop $P$. We know that in the deconfined phase
$P$ clusters around the values $\e^{\im \theta_P}$, with $\theta_P =
0$ (real sector) and $\theta_P = \pm2\pi/3$ (complex sector)
\cite{Z3}. Because the fermionic action does not share the $Z_3$
symmetry of the gauge action, the localization of the eigenmodes of
the Dirac operator can depend on the Polyakov loop sectors. Indeed one
finds that the zero mode $\psi_0$ of the caloron is more localized in
the real sector than in the complex sector, as can be seen from
\cite{own,cal}
\begin{equation}
\left| \psi_0 \right|^2 \propto \exp[ -2 (\pi - |\theta_P| ) \,
  r T ]/r^2.
\end{equation}
Next we will search for caloron-like configurations concentrating on
the fermionic properties of the caloron described above.

\begin{figure*}[htp]
\centerline{ \epsfig{file=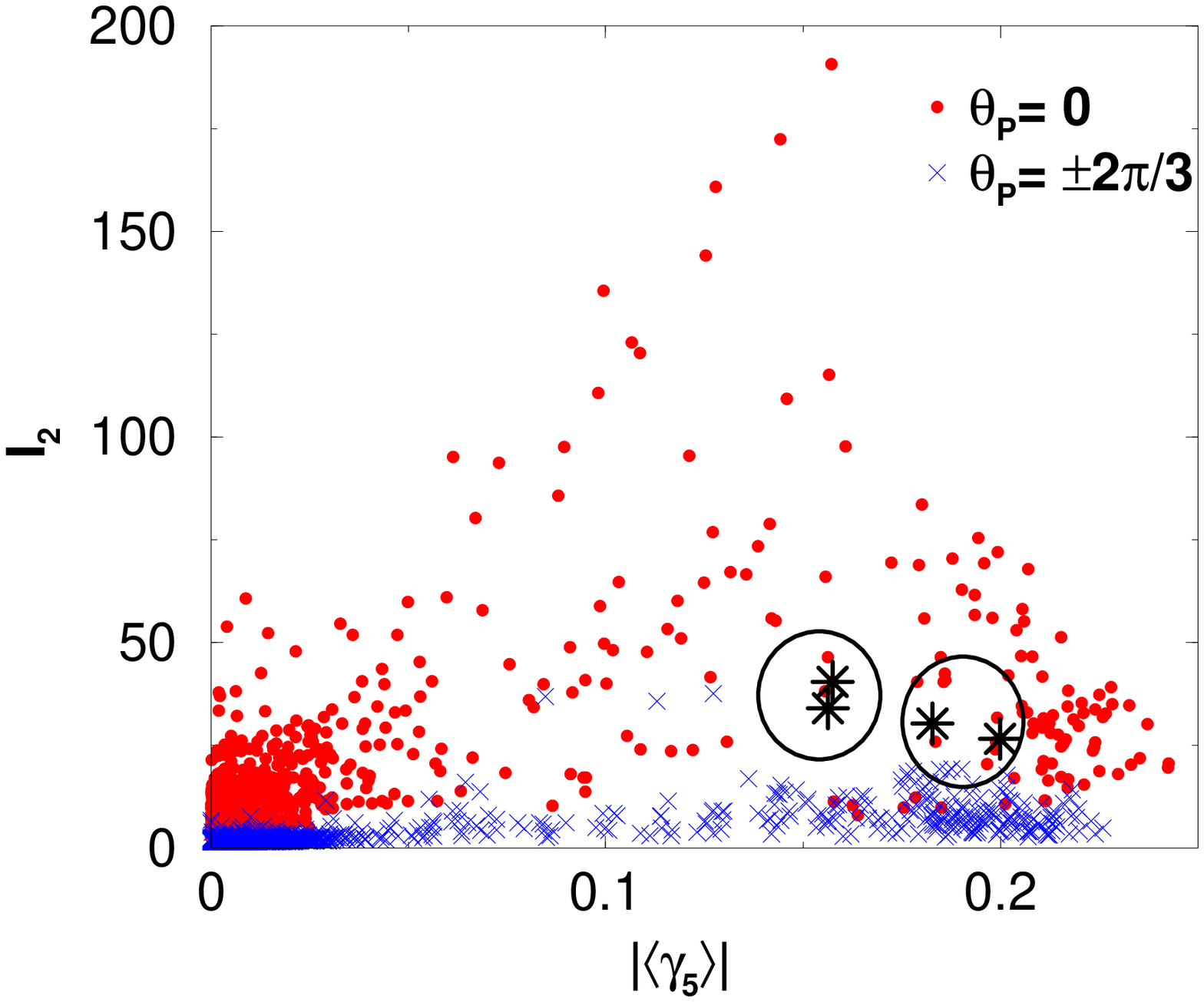, width = 6.2cm} \hspace{5mm}
  \epsfig{file=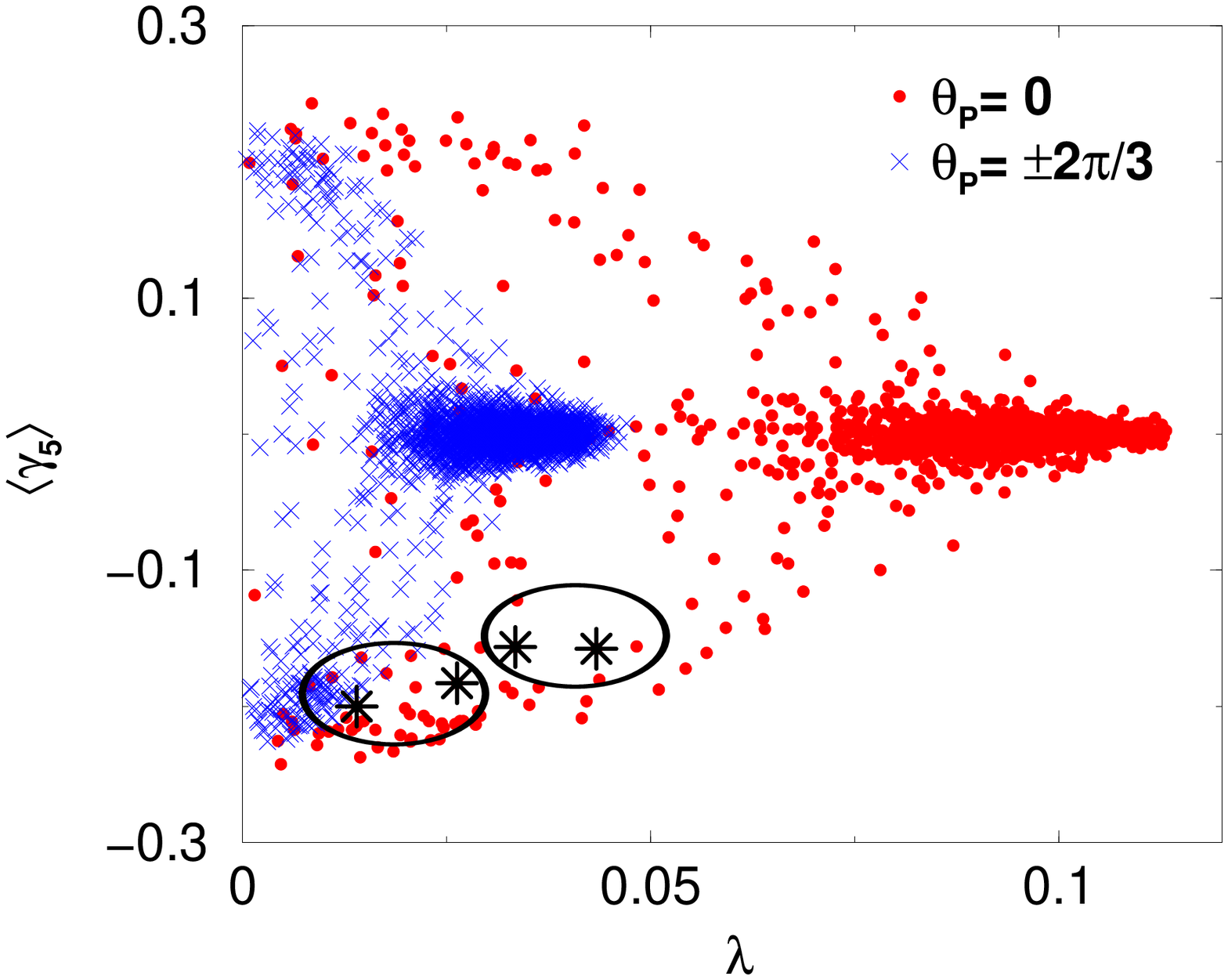, width = 6.2cm}}
\vspace{-9mm}
\caption{\label{fig1}Expectation value of $\gamma_5$ vs.\ eigenvalue
$\lambda$ (left) and inverse participation ratio vs.\ $\left| \left<
\gamma_5 \right> \right|$ (right). Results for the lowest 10
eigenmodes of $\approx 300$ configurations at $\beta=8.45$ on a
$20^3\times 6$ lattice are shown.}
\end{figure*}

\section{LATTICE SETUP}
We work in the quenched approximation with the L\"uscher-Weisz gauge
action
\begin{eqnarray}
S_g[U] & = & \textstyle \beta_1 \sum_{pl} \frac{1}{3} \Re \; \Tr \; [ 1 - U_{pl} ] +
 \nonumber \\
 &  & \textstyle \beta_2 \sum_{rt} \frac{1}{3} \Re \; \Tr \; [ 1 - U_{rt} ] +
\nonumber \\ 
 &  & \textstyle \beta_3 \sum_{pg} \frac{1}{3} \Re \; \Tr \; [ 1 - U_{pg} ],
\end{eqnarray}
where $\sum_{pl}$, $\sum_{rt}$ and $\sum_{pg}$ means a summation over
all plaquettes, $2 \times 1$ rectangles and parallelograms,
respectively. The coefficients $\beta_2$ and $\beta_3$ are computed
from $\beta_1$ via tadpole improved pertubation theory. Further we use 
the staggered Dirac operator
\begin{displaymath}
 D=  \sum_{\mu=1}^4  \frac{1}{2a} \alpha_{\mu}(x) \, 
 \left[\delta_{y,x+\hat \mu} U_\mu(x)
 - \delta_{y,x-\hat \mu}U_\mu^\dagger(y) \right] \:,
\end{displaymath} 
where $\alpha_{\mu}(x) = (-1)^{x_1 + \ldots + x_{\mu-1}}$. The
eigenvalues of $D$ come in pairs of $\pm \rm{i} \lambda$ with
$\lambda$ real, so we can restrict ourselves to positive $\lambda$ in
the following. In the continuum limit this action corresponds to four
quark flavors. From now on we set $a$ to 1.

We have calculated the lowest eigenvalues and eigenvectors for lattice
sizes $12^3 \times 6$, $16^3 \times 6$ and $20^3 \times 6$ below the
deconfinement phase transition at $\beta_1 = 8.10$
($\beta_\mathrm{Wilson} \approx 5.8$
\cite{gat}) and above at $\beta_1 = 8.45$ ($\beta_\mathrm{Wilson}
\approx 6.0$ \cite{gat}) with the Arnoldi method~\cite{Arnoldi}. We
will present data for the largest lattice at $\beta=8.45$.
\begin{figure*}[htp]
  \centerline{ 
    \epsfig{file=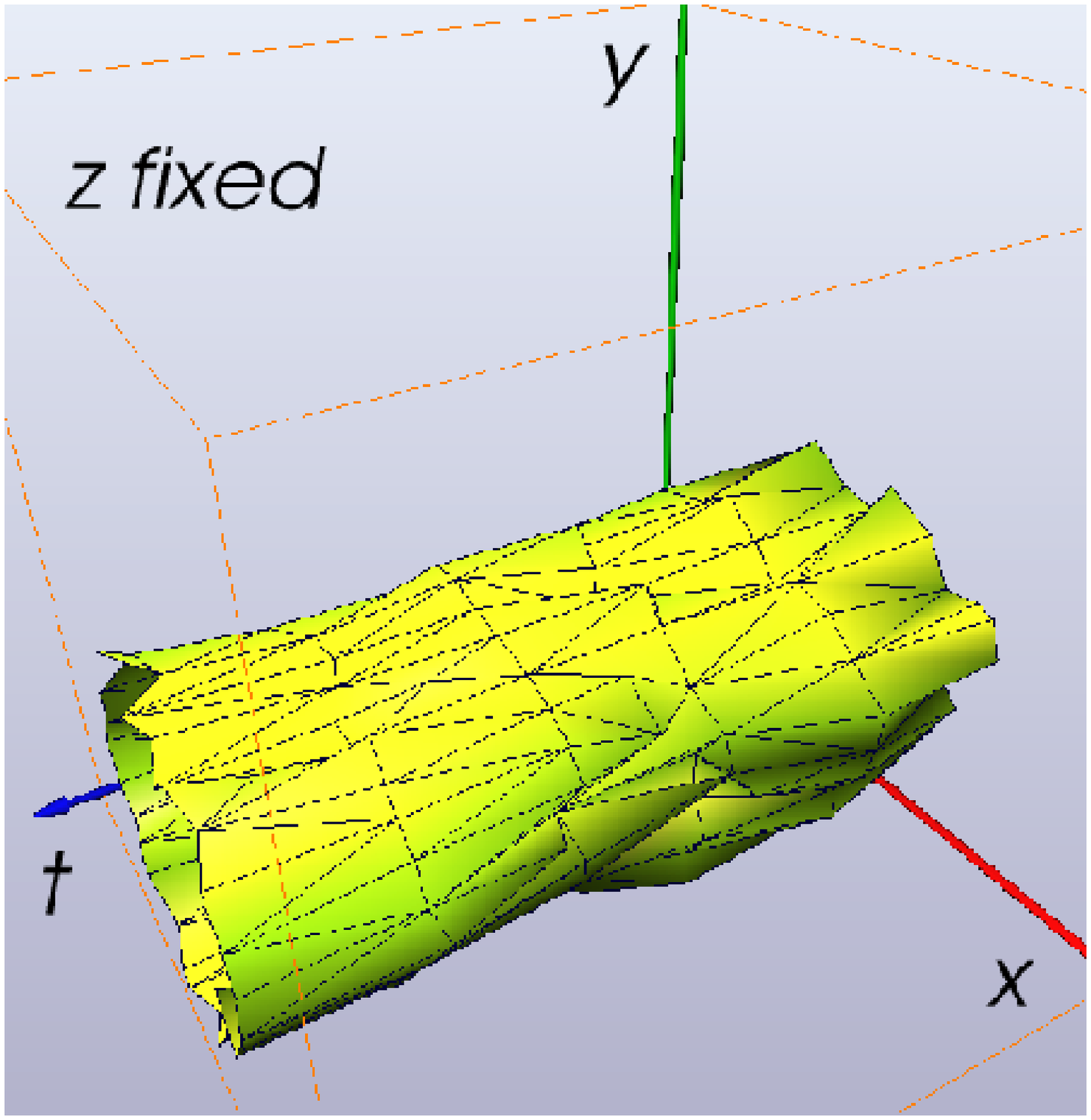, width = 4.8cm} 
    \epsfig{file=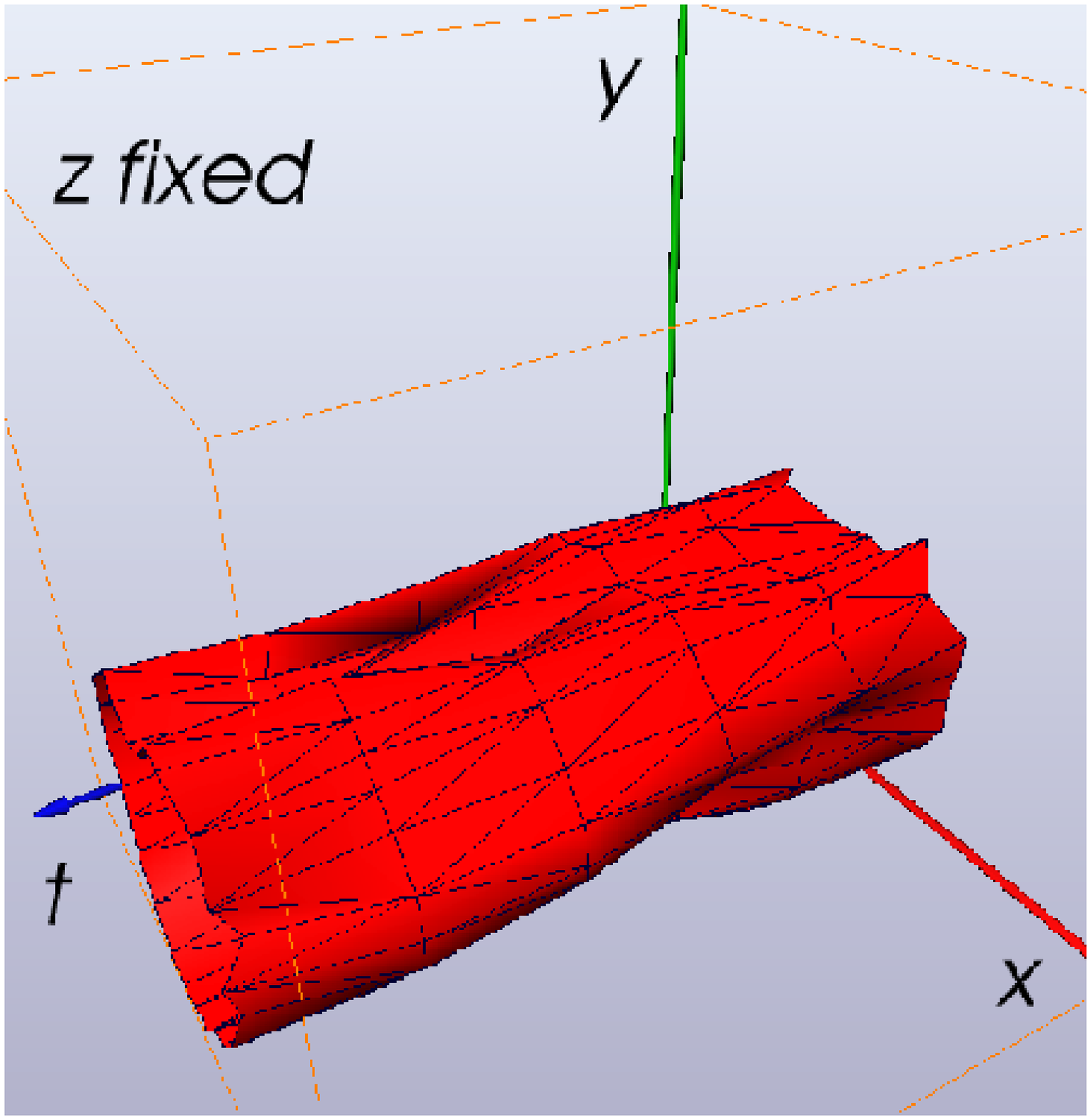, width = 4.8cm}
    \epsfig{file=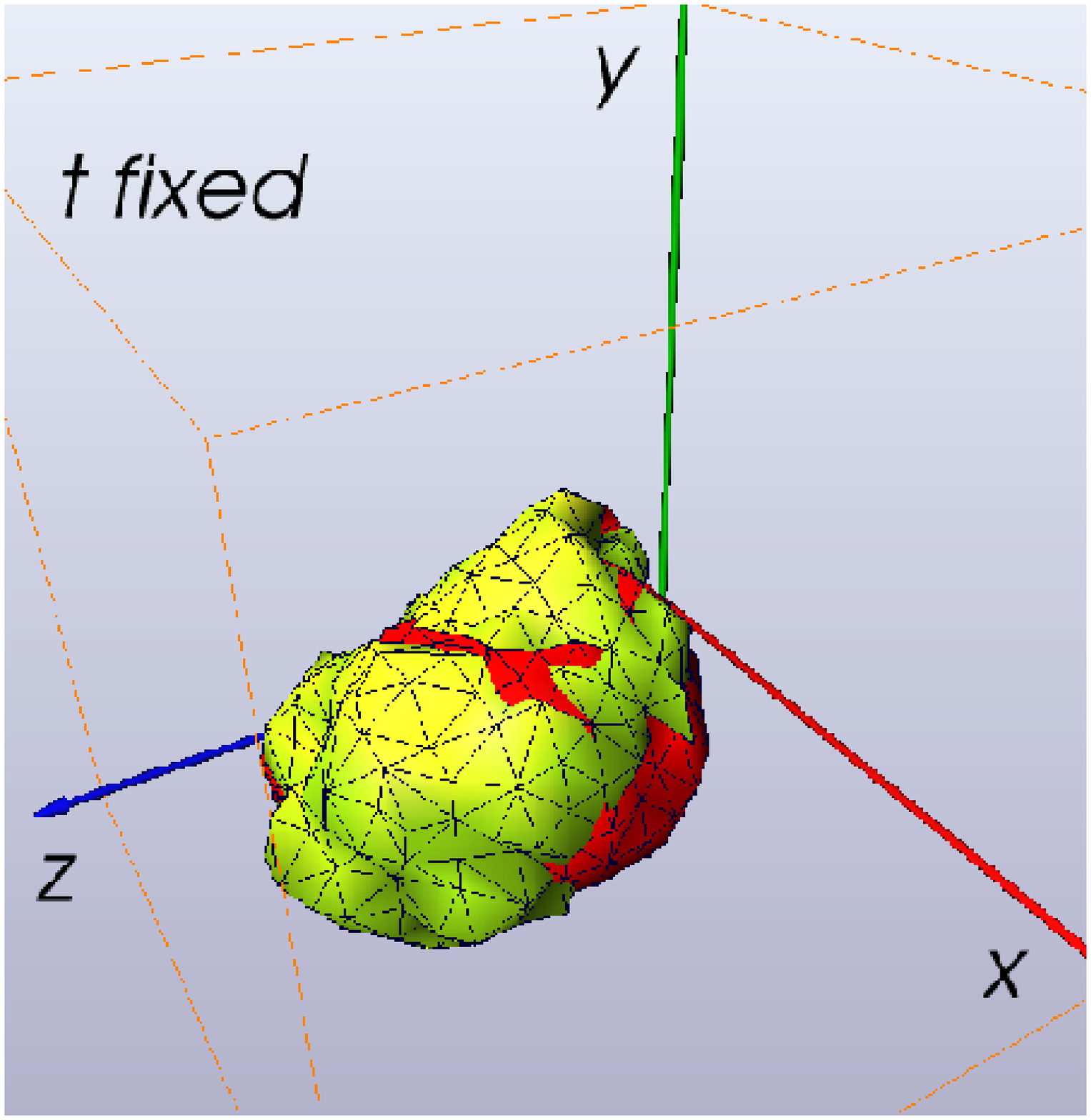, width = 4.8cm}}
\vspace{-5mm}
\caption{\label{fig2}Isosurfaces of $p_{\lambda}$ (light grey) and
$p_{\lambda}^5$ (dark grey) for one of the marked eigenmodes of
Fig. \ref{fig1}. The left and middle plots show them separately at
fixed $z$, in the right plot they are overlaid at fixed $t$.}
\end{figure*}
\section{CALORONS ON THE LATTICE}
When searching for calorons on the lattice with the help of their
fermionic properties, the zero mode of the Dirac operator plays the
central role. Unfortunately the staggered Dirac operator has no exact
zero modes, but, as we can see clearly in the left plot of
Fig.~\ref{fig1}, there are for both Polyakov loop sectors low-lying
eigenmodes with a high value of $\left| \left< \gamma_5 \right>
\right|$ and eigenmodes in the bulk of the spectrum with values of $\left<
\gamma_5 \right>$ around zero. We find that the low-lying
eigenvalues are quasi 4-fold degenerate in the sense that the number
of low-lying eigenmodes is almost always 4, 8, 12,~\ldots (In
Fig.~\ref{fig1} this is illustrated by the encircled eigenmodes, which
belong to the same gauge field configuration. Note that the total
number of low-lying eigenmodes of that configuration is 8.) So one
would expect that those low-lying eigenmodes correspond to exact zero
modes with definite handedness in the continuum.

Let us now look at the localization properties of these approximate
zero modes. We define a gauge-invariant measure of the localization of
a quark eigenmode $\psi^\alpha_{\lambda}(x)$ ($\lambda$ is the Dirac
eigenvalue, $\alpha$ a color index), the inverse participation ratio
\begin{equation}
  I_2 \equiv V \frac{\sum_x p_{\lambda}(x)^2 }
  {\left[\sum_x p_{\lambda}(x)\right]^2}\:,
  \label{Idef} 
\end{equation} 
where $V$ is the number of lattice sites and $p_{\lambda}(x)$ is the
gauge-invariant probability density $p_{\lambda}(x) = \sum_{\alpha =
1}^{N_c} \left| \psi^\alpha_{\lambda}(x) \right|^2$. For a completely
delocalized state (all $p_\lambda(x)$ the same) one finds $I_2 =1$,
whereas a state localized on a single lattice site (only one non-zero
$p_\lambda(x)$) would have $I_2 = V$. 
In the right part of Fig.~\ref{fig1} we plot $I_2$ vs.\ $\left| \left<
\gamma_5 \right> \right|$. We see that the modes in the real sector
are, in general, more localized than those in the complex sector. In
both Polyakov loop sectors the approximate zero modes have a greater
value of $I_2$ than the modes in the bulk spectrum, which means that
they are more localized. These are exactly the properties we expect for
caloron configurations.

Finally we consider the localization pattern. In Fig.~\ref{fig2}
isosurfaces of $p_\lambda$ and the local chirality $p_{\lambda}^5(x) =
\sum_{\alpha = 1}^{N_c} {\psi^\alpha_{\lambda}(x)}^\ast \gamma_5
\psi^\alpha_{\lambda}(x)$ are plotted at fixed $t$ and at fixed $z$
for one of the encircled eigenmodes. This (typical) mode is clearly
localized in space but not in time and the isosurfaces for $p_\lambda$
and $p_{\lambda}^5$ are very similar. This is again as is expected for
caloron-like configurations.

\section{CONCLUSIONS}
Studying the localization and chirality properties of low-lying quark
eigenstates in quenched lattice QCD, we could characterize
semiclassical properties of the gauge field configurations without any
cooling. For temperatures above $T_c$ we found isolated modes with
definite handedness. They are localized in space but not in time with
characteristic differences between the Polyakov loop sectors. Thus
they show essential properties of quark states associated with
calorons.
\\\\
{\bf Acknowledgements:} We thank the LRZ in Munich for computer time
on the Hitachi SR8000. Support from BMBF and DFG is acknowledged.

\end{document}